\documentclass[]{aa}
\usepackage{epsfig}

\def\gsim{\vcenter{\hbox{$>$}\offinterlineskip\hbox{$\sim$}}}

\begin{document}
\title{Mid-infrared observations of young stellar objects\\
in the vicinity of $\sigma$ Orionis
\thanks{Based on observations collected at the European Southern Observatory,
Chile (ESO N$^\circ$ 70.C-0623)}
}
\author{Joana M. Oliveira and Jacco Th. van Loon}
\institute{Astrophysics Group, School of Chemistry \& Physics, Keele
           University, Staffordshire ST5 5BG, United Kingdom}
\offprints{\email{joana@astro.keele.ac.uk}}
\date{Received date; accepted date}
\titlerunning{Mid-IR observations of YSOs in the vicinity of $\sigma$ Ori}
\authorrunning{Oliveira \& van Loon}
\abstract{
We present new mid-infrared observations of objects in the vicinity of the O-star $\sigma$\,Orionis, obtained with TIMMI-2 at ESO. By constraining their near- and mid-infrared spectral energy distributions, we established the nature of previously known IRAS sources and identified new mid-infrared sources as young stellar objects with circumstellar disks, likely massive members of the $\sigma$\,Ori cluster. For two of these objects we have obtained spectroscopy in the 8--13\,$\mu$m range in order to investigate the chemistry of the dust grains. TX\,Ori exhibits a typical silicate emission feature at 10\,$\mu$m, with a feature at about 11.2\,$\mu$m that we identify as due to crystalline olivine. The IRAS\,05358$-$0238 spectrum is very unusual, with a weak silicate feature and structure in the range 10--12\,$\mu$m that may be explained as due to self-absorbed forsterite. We also provide the first evidence for the presence of circumstellar disks in the jet sources Haro\,5-39/HH\,447, V510\,Ori/HH\,444 and V603\,Ori/HH\,445.
}
\maketitle

\section{Introduction}

Disk-like structures seem to be ubiquitous during the formation and early evolution of low-mass stars. Disks are also the birthplace of planetary bodies. A particularly challenging problem is how to reconcile the relatively long timescales for planet formation ($>$\,10\,Myr, Bodenheimer et al.\ \cite{bodenheimer00}) and the rather quick destruction of circumstellar disks (e.g.\ Haisch et al.\ \cite{haisch01a}). Amongst the many (yet) unanswered disk-related questions are: what are the timescales of disk dissipation; how does the chemical and physical evolution of dust proceed from small interstellar dust grains through pebble-sized particles to larger bodies? And how does the local physical environment (in particular in OB associations) influence these processes?

Circumstellar disks have traditionally been identified from near-infrared (near-IR) colours ($JHK$). Recently, several L-band surveys of young stellar populations proved that the K-band excess is a rather incomplete and unreliable disk indicator. Furthermore, theoretical work on the IR signatures of circumstellar disks suggests that L-band observations can detect disks even for very low disk masses (Wood et al.\ \cite{wood02}). Thus L-band surveys are very efficient in detecting circumstellar material and are thought to be largely complete for very young clusters --- Haisch et al\ (\cite{haisch01b}) found this to be case for the embedded cluster NGC\,2024. However, for older clusters, detection in the L-band might become more difficult if disk evolution leads to the removal of the hotter circumstellar dust component. Furthermore these surveys still do not provide enough information on the geometry of the system (e.g.\ protostellar object versus young stellar object with disk) and they provide little insight into the properties of the circumstellar material.

Mid-IR observations can unequivocally identify circumstellar material, constrain the spectral energy distribution and the physical parameters of the observed system and, through spectroscopy, allow the identification of the chemical and mineral species present in the dust grains. However such observations are technically challenging and relatively few clusters and associations have been surveyed in the N-band (e.g.\ Taurus-Auriga, Kenyon \& Hartmann \cite{kenyon95}; $\rho$\,Ophiuchi cloud, Green et al.\ \cite{green94}; NGC\,2024, Haisch et al.\ \cite{haisch01b}; NGC\,3603, N\"{u}rnberger \& Stanke \cite{nurnberger03}), and mid-IR spectroscopy (around 10\,$\mu$m) has mostly concentrated on the more massive Herbig Ae/Be objects (Bouwman et al.\ \cite{bouwman01}) and objects in the Taurus-Auriga and $\rho$\,Ophiuchi complexes (Hanner et al.\ \cite{hanner95,hanner98}).

$\sigma$\,Orionis is a Trapezium-like system with an O9.5\,V primary. The population of low-mass stars spatially clustered around this system was discovered as bright X-ray sources in ROSAT images (Wolk \cite{wolk96}; Walter et al.\, \cite{walter97}). A recent L$'$-band survey of low-mass $\sigma$\,Orionis cluster members has revealed that $\sim$\,46\% of these objects have circumstellar disks, at a cluster age of 3$-$5\,Myr (Oliveira et al.\ \cite{oliveira04}). A mid-infrared source has been discovered very close to $\sigma$ Orionis itself, apparently a proto-planetary disk being dispersed by the intense ultraviolet radiation from this massive star (van Loon \& Oliveira \cite{loon03}). A few IRAS sources were known in the vicinity of $\sigma$\,Ori. In this paper we describe new mid-IR imaging observations within an area around $\sigma$\,Ori, aimed at revealing the nature of the known mid-IR sources and detecting mid-IR emission from other dusty pre-main-sequence (PMS) stars. For a few of these objects we obtained 8$-$13\,$\mu$m spectra in order to determine the composition of the circumstellar dust.

\section{Observations}

\subsection{Target selection}

\begin{figure}[tb]
\resizebox{\hsize}{!}{\includegraphics{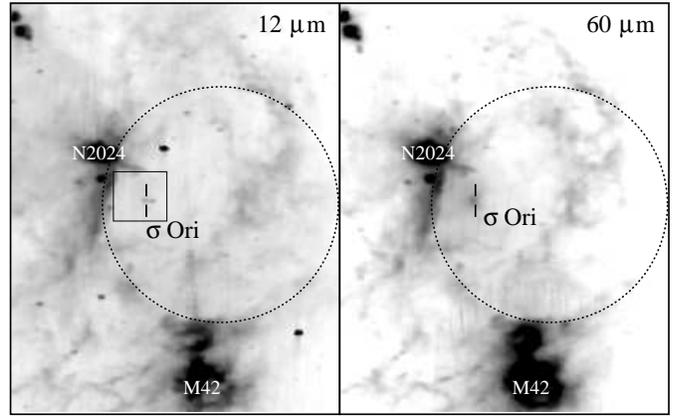}}
\caption{IRAS 12 and 60\,$\mu$m images of part of the Orion molecular clouds, showing $\sigma$\,Orionis and sites of recent (embedded) star formation NGC\,2024 and M\,42 (Orion Nebula). North is to the top and East to the left. The box represents the area enlarged in Fig.\,\ref{f2}.}
\label{f1}
\end{figure}

\begin{figure}[tb]
\resizebox{\hsize}{!}{\includegraphics{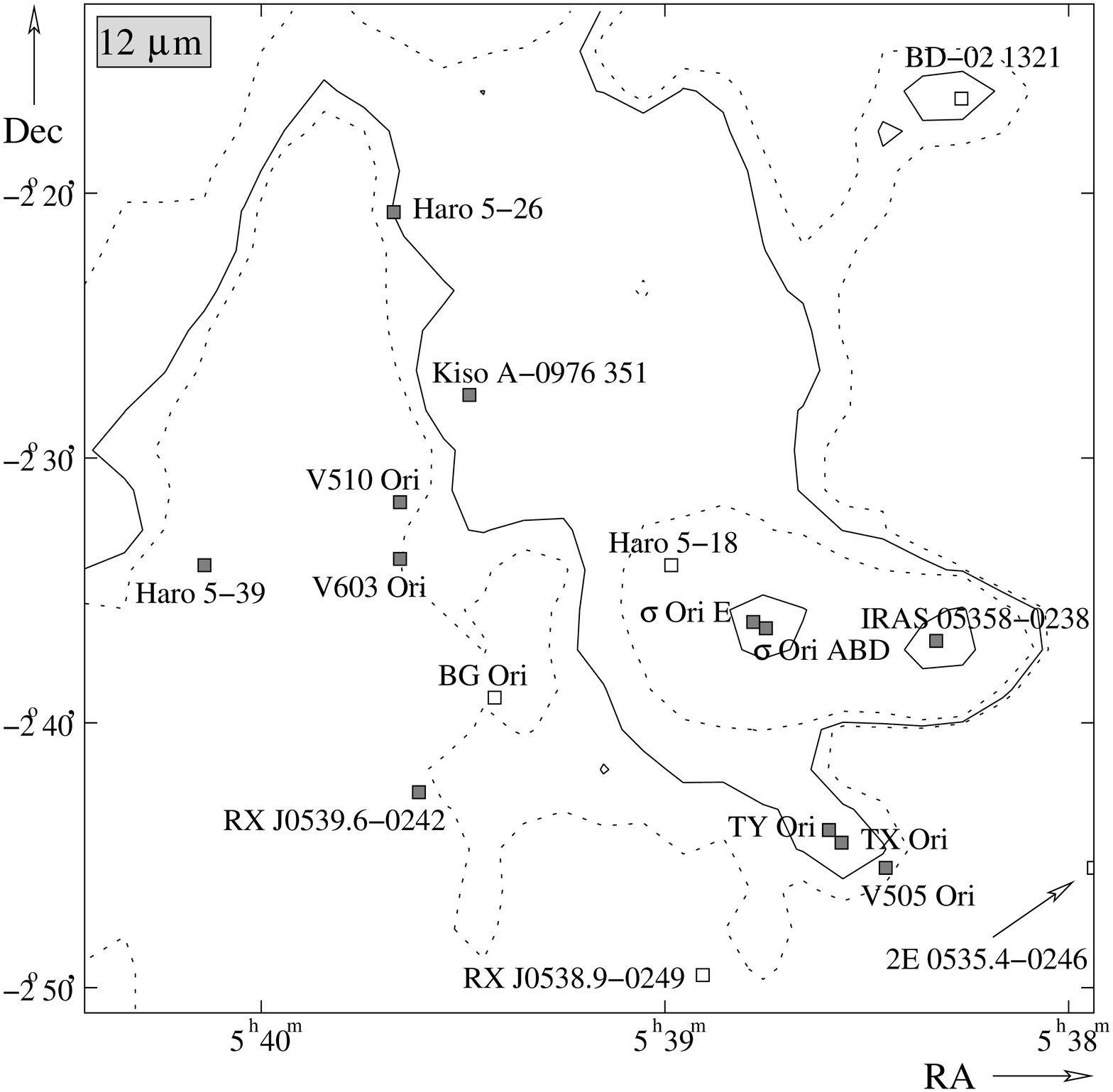}}
\caption{IRAS 12\,$\mu$m contour map of the immediate neighbourhood of $\sigma$\,Ori. The filled squares are objects we have observed with TIMMI-2.}
\label{f2}
\end{figure}

An IRAS 12\,$\mu$m image (Fig.\,\ref{f1}) of the star formation sites around and below the Orion belt shows that the area near $\sigma$\,Ori itself is largely devoid of warm and/or dense dust clouds, although, curiously enough, this ``cavity'' seems to be filled with either cold dust or ionized gas shining at 60\,$\mu$m. The contour map of 12\,$\mu$m emission in Fig.\,\ref{f2} zooms into the $38^\prime\times38^\prime$ area closer to the $\sigma$\,Ori system. Few mid-infrared sources were detected with IRAS in this region, due in part to the poor resolution and relatively low sensitivity. The IRAS Point Source Catalogue (PSC) includes a bright source centred approximately on $\sigma$ Ori AB (IRAS\,05362$-$0237) and a second bright source to the W, IRAS\,05358$-$0238. Only $\sim40^{\prime\prime}$ away from $\sigma$\,Ori AB, $\sigma$\,Ori\,E is another hot star for which excess emission has been detected at 3.5 and 5\,$\mu$m (Groote \& Hunger \cite{groote82}), but its IRAS flux densities could not be derived due to the proximity of $\sigma$\,Ori AB. Nearby PMS stars detected by IRAS (Weaver \& Jones \cite{weaver92}) are TX\,Ori and TY\,Ori (that appear blended) and V510\,Ori. Another, more distant mid-IR point source, IRAS\,05357$-$0217 is associated with the F8-type star BD$-$02\,1321, of which the IR emission has been analysed previously (e.g\, Garc\'{\i}a-Lario et al.\, \cite{garcia90}). There are a number of other PMS objects in the field, that could have dust associated with them but which IRAS could not detect or distinguish from the few brighter sources. Indeed, the IRAS 12\,$\mu$m contour map suggests that there might well be unresolved emission to the N-NE and possibly also to the SE of $\sigma$\,Ori.

\begin{figure}[tb]
\resizebox{\hsize}{!}{\includegraphics{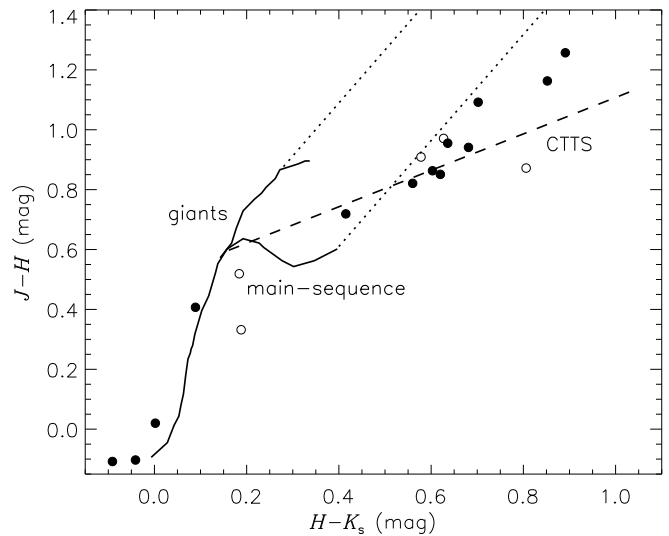}}
\caption{$JHK_{\rm s}$ colour-colour diagram of targets of the TIMMI-2 observations, filled symbols are the objects effectively observed. The full lines are the loci of main-sequence and giant stars (Bessell \& Brett \cite{bessell88}), the dotted lines are reddening bands (Bessell \& Brett \cite{bessell88}; Rieke \& Lebofsky \cite{rieke85}), the dashed line is the locus of classical T\,Tauri stars (Meyer et al.\, \cite{meyer97}), converted to the 2MASS photometric system (Carpenter \cite{carpenter01}). Objects with colours to the right of the reddening band cannot be explained simply by photospheric emission from the star and suggest the presence of a circumstellar disk.}
\label{f3}
\end{figure}

At the time of the observations described here, the circumstellar disk population around $\sigma$\,Ori remained largely unsurveyed. To compile a sample of candidate circumstellar disk objects, we cross-correlated catalogues that included suspected PMS objects (Orion variables, emission-line stars, X-ray sources, etcetera) with the Two Micron All-Sky Survey (2MASS) Point Source Catalog. In the $(J-H)$ versus $(H-K_{\rm s})$ colour-colour diagram, stars with $(H-K_{\rm s})>0.55$\,mag are located to the right of the reddening band (see Fig.\,\ref{f3}) and thus cannot be explained as purely stellar photospheric emission, being therefore candidates for objects with circumstellar disks. We note, however, that the absence of $(H-K_{\rm s})$ excess does not necessarily imply the absence of a circumstellar disk (e.g.\, Oliveira et al.\ \cite{oliveira04}). V510\,Ori, Haro\,5-39, V603\,Ori and BG\,Ori are the sources of the well-known irradiated protostellar jets HH\,444, HH\,447, HH\,445 and HH\,446 respectively (Reipurth et al.\, \cite{reipurth98}). All these objects mingle with the spectroscopically confirmed $\sigma$\,Ori cluster members; the observational evidence of young age makes them likely cluster members, even though cluster membership cannot be ascertained at this stage. Zapatero Osorio et al.\, (\cite{osorio02}) has confirmed membership for V505\,Ori.

\begin{table*}[ht]
\caption{Objects observed with TIMMI-2, with their most common names,
2MASS coordinates (except for $\sigma$\,Ori\,IRS1, coordinates from van Loon \& Oliveira \cite{loon03}), and spectral type.}
\label{t1}
\begin{tabular}{llllll}
\hline\hline
Object name & Alternative name(s) & \multicolumn{2}{c}{Coordinates (J2000)}&SpT & Remarks \\
            &                     & RA($^{\rm h}$ $^{\rm m}$ $^{\rm s}$) &Dec($^{\rm d}$ $^{\rm m}$ $^{\rm s}$)&  & \\
\hline
$\sigma$\,Ori\,AB  &ADS\,4241\,AB, HD\,37468\,AB, 48\,Ori           &5 38 44.8&$-$2 36 00&O9.5V&\\
$\sigma$\,Ori\,D   &ADS\,4241\,D, HD\,37468\,D                      &5 38 45.6&$-$2 35 59&B2V  &\\
$\sigma$\,Ori\,E   &ADS\,4241\,E, HD\,37479, V1030\,Ori             &5 38 47.2&$-$2 35 41&B2V  &Helium-rich\\                                   
$\sigma$\,Ori\,IRS1&IRAS\,05362$-$0237                              &5 38 44.9&$-$2 35 57&     &proplyd?\\
RX\,J0539.6$-$0242 &                                                &5 39 36.5&$-$2 42 17&K0   &T\,Tauri, double\\
Haro\,5-26         &HV\,1179, Kiso\,A-0904\,91, RV\,Ori             &5 39 40.2&$-$2 20 48&     &\\
Haro\,5-39         &Kiso\,A-0976\,364, Kiso\,A-0904\,100, V608\,Ori?&5 40 08.9&$-$2 33 34&     &source of HH\,447\\
IRAS\,05358$-$0238 &                                                &5 38 19.8&$-$2 36 39&     &\\
Kiso\,A-0976\,351  &                                                &5 39 29.4&$-$2 27 21&     &Double\\
TX\,Ori            &Haro\,5-12, HV\,797, IRAS\,05360$-$0245         &5 38 33.7&$-$2 44 14&K4   &T\,Tauri\\                                       
TY\,Ori            &HV\,798                                         &5 38 35.9&$-$2 43 51&K3   &T\,Tauri\\
V505\,Ori          &                                                &5 38 27.3&$-$2 45 10&K7   &\\
V510\,Ori          &Haro\,5-27, HV\,1180, Kiso\,A-0976\,356         &5 39 39.8&$-$2 31 22&     &source of HH\,444\\
V603\,Ori          &                                                &5 39 39.8&$-$2 33 16&     &source of HH\,445?\\
\hline
\end{tabular}
\end{table*}

Our sample thus consists of the 6 known IRAS point sources plus 11 objects that had near-IR excess and/or were previously suspected to be classical T\,Tauri stars (CTTS); we observed 12 of these targets, which are listed in Table\,\ref{t1}. We resolve the different components of the $\sigma$\,Ori system: the components D and E are analysed here, whilst the spectral energy distribution of $\sigma$\,Ori\,AB and the unexpected discovery of the fascinating new mid-IR source $\sigma$\,Ori\,IRS1 have already been reported (van Loon \& Oliveira \cite{loon03}). Oliveira et al.\ (\cite{oliveira04}) have since identified a population of cluster members with $L'$-band excesses, including 6 of the late-type objects in our sample. Table\,\ref{t2} lists the 2MASS $J$, $H$ and $K_{\rm s}$ measurements, L- or L$'$-band measurements when available (Castor \& Simon \cite{castor83}; Oliveira et al.\, \cite{oliveira04}), and our new N1- and Q1-band measurements. The IRAS flux densities of $\sigma$\,Ori\,IRS1, IRAS\,05358$-$0238, TX\,Ori and V510\,Ori have been re-measured from the original IRAS scans (see Appendix\,A).

\begin{table*}[ht]
\caption{Infrared photometry for the objects observed with TIMMI-2. The J, H
and K-band magnitudes are from 2MASS except when also observed with UKIRT in
the L$^\prime$-band in which case the K and L$^\prime$-band magnitudes are
from Oliveira et al.\ (2003). The exception is $\sigma$\,Ori\,AB for which the
JHKL photometry is taken from Castor \& Simon (1983). Mid-IR observations of $\sigma$\,Ori\,IRS1 are from van Loon \& Oliveira (\cite{loon03}). The N1- and Q1-band flux densities and N-band spectra are from this work.}
\label{t2}
\begin{tabular}{lccccccc}
\hline\hline
Object&J&H&K&L$^\prime$&N1&\hspace*{5mm}Q1&N-band\\
      &(mag$\pm\sigma$)&(mag$\pm\sigma$)&(mag$\pm\sigma$)&(mag$\pm\sigma$)&(Jy$\pm\sigma$)&\hspace*{5mm}(Jy$\pm\sigma$)&spectrum\\
\hline
$\sigma$\,Ori\,AB  &        4.301$\pm$0.008&	    4.409$\pm$0.007&        4.500$\pm$0.008&4.544$\pm$0.017&0.617$\pm$0.031&&yes\\
$\sigma$\,Ori\,D   &        7.116$\pm$0.029&	    7.219$\pm$0.027&        7.260$\pm$0.021&               &0.051$\pm$0.006&&\\
$\sigma$\,Ori\,E   &        6.974$\pm$0.025&	    6.954$\pm$0.031&        6.952$\pm$0.029&               &0.076$\pm$0.010&&\\
$\sigma$\,Ori\,IRS1&                       &                       &                       &               &0.573$\pm$0.029&2.38$\pm$\rlap{0.24}&yes\\
RX\,J0539.6$-$0242 &        8.462$\pm$0.027&	    8.055$\pm$0.040&        7.966$\pm$0.004&8.020$\pm$0.035&0.037$\pm$0.004&&\\
Haro\,5-26         &\llap{1}1.495$\pm$0.024&\llap{1}0.632$\pm$0.024&\llap{1}0.029$\pm$0.021&               &\llap{$<$}0.012		&&\\
Haro\,5-39         &\llap{1}1.501$\pm$0.026&\llap{1}0.546$\pm$0.023&        9.812$\pm$0.006&8.762$\pm$0.062&0.072$\pm$0.008&&\\
IRAS\,05358$-$0238 &        9.410$\pm$0.028&	    8.318$\pm$0.055&        7.616$\pm$0.018&               &5.467$\pm$0.273&3.69$\pm$\rlap{0.37}&yes\\
Kiso\,A-0976\,351  &\llap{1}2.843$\pm$0.030&\llap{1}2.022$\pm$0.026&\llap{1}1.462$\pm$0.026&               &\llap{$<$}0.017              &&\\
TX Ori             &\llap{1}0.131$\pm$0.026&	    9.280$\pm$0.024&        8.600$\pm$0.002&7.715$\pm$0.024&0.291$\pm$0.015&0.53$\pm$\rlap{0.11}&yes\\
TY Ori             &\llap{1}0.445$\pm$0.027&	    9.726$\pm$0.024&        9.311$\pm$0.028&               &0.054$\pm$0.004&&\\
V505\,Ori          &\llap{1}1.955$\pm$0.028&\llap{1}0.792$\pm$0.026&        9.729$\pm$0.006&8.642$\pm$0.044&0.074$\pm$0.005&&\\
V510\,Ori          &\llap{1}1.842$\pm$0.030&\llap{1}0.901$\pm$0.023&        9.831$\pm$0.006&8.483$\pm$0.069&0.167$\pm$0.009&&\\
V603\,Ori          &\llap{1}2.218$\pm$0.026&\llap{1}0.961$\pm$0.024&\llap{1}0.387$\pm$0.009&9.088$\pm$0.060&0.050$\pm$0.005&&\\
\hline
\end{tabular}
\end{table*}

\subsection{Technical details}

The mid-IR imager and spectrograph TIMMI-2 at the ESO 3.6m telescope at La
Silla, Chile, was used on the nights of 15$-$18 December 2002. Images of the
targets were obtained through the N1-band filter ($\lambda_0=8.6$ $\mu$m,
$\Delta\lambda=1.2$\,$\mu$m for Full-Width at Half Maximum (FWHM) and
$\Delta\lambda=1.7$\,$\mu$m between blue and red cut-off). On the first two nights, images of the brighter IR sources $\sigma$\,Ori, IRAS\,05358$-$0238 and TX\,Ori were also obtained through the Q1-band filter ($\lambda_0=17.75$ $\mu$m, $\Delta\lambda=0.8$\,$\mu$m for FWHM and $\Delta\lambda=1.4$\,$\mu$m between blue and red cut-off). These narrow-band filters provide better sensitivity and less risk of saturating the array by the thermal background. As the N1-band filter is not centred at the peak of the silicate dust feature at 10\,$\mu$m it more closely represents the brightness of the continuum (either stellar or circumstellar), but the Q1-band filter is quite sensitive to the strength of the silicate feature at 18\,$\mu$m. The N1-band may be compared with, for instance, the often used ISO LW2 filter or the (much broader) MSX band A.

The pixel scale was $0.2^{\prime\prime}$ pixel$^{-1}$, resulting in a $64^{\prime\prime}\times48^{\prime\prime}$ (RA$\times$Dec) field-of-view. We used a chop throw of $10^{\prime\prime}$ in the N-S direction and a nod offset of $10^{\prime\prime}$ in the E-W direction. The frames resulting from the on-line data reduction pipeline were shift-added within the ESO software package {\sc midas} to create stellar images with a FWHM of 0.7$-$1.0$^{\prime\prime}$. Photometry was performed using a circular software aperture with a $2^{\prime\prime}$ diameter and it was then calibrated against HD\,4128 ($N1=69.61$\,Jy, $Q1=16.97$\,Jy) and HD\,32887 ($N1=65.61$\,Jy, $Q1=17.56$\,Jy).

\begin{figure*}[ht]
\includegraphics[width=17cm]{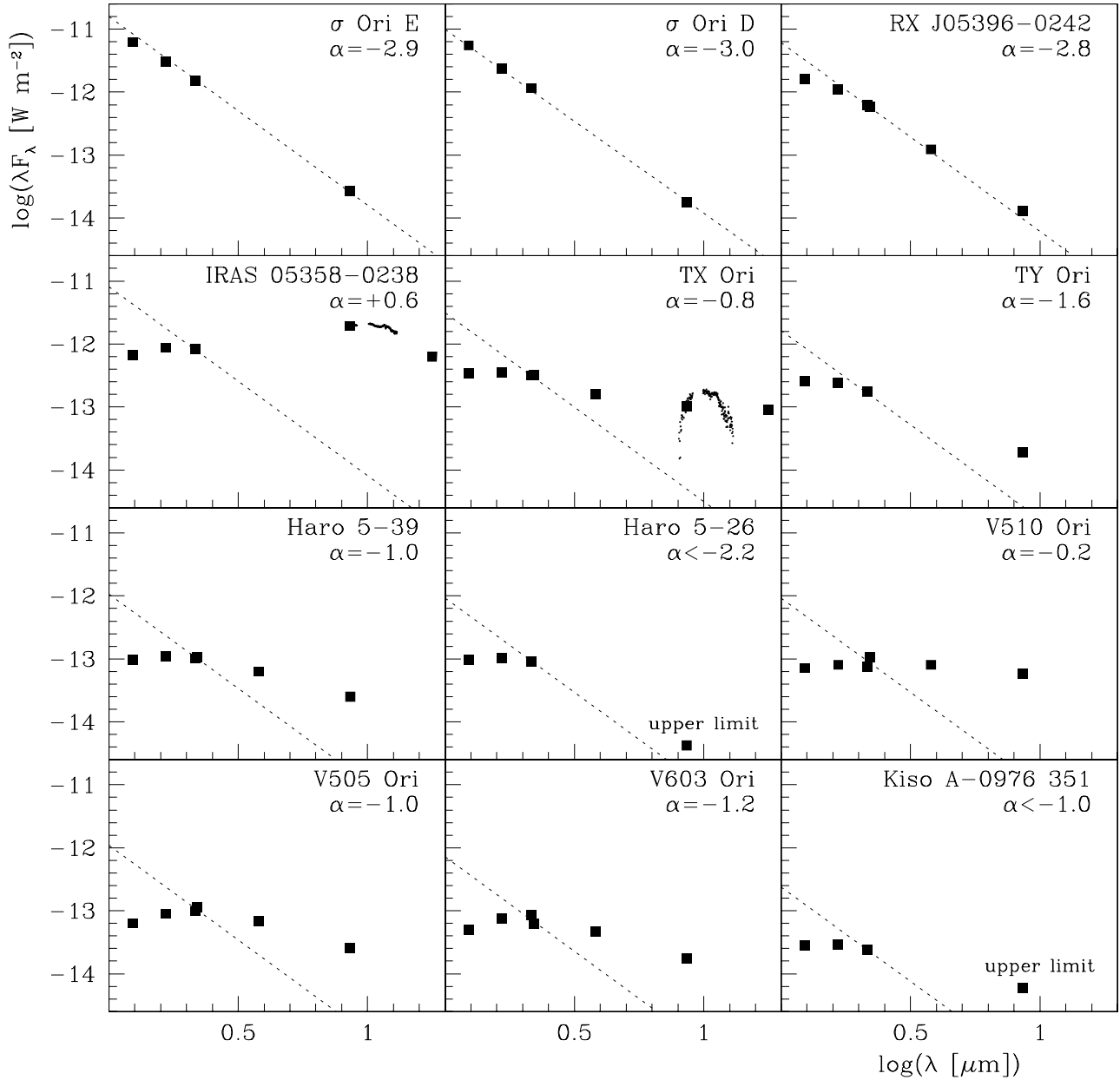}
\caption{Infrared spectral energy distributions for all targets except
$\sigma$\,Orionis and $\sigma$\,Ori IRS1, ordered in luminosity (from left to
right and from top to bottom). The spectral index is defined as
$\alpha=(F_{\rm N1}-F_{\rm K})/(\lambda_{\rm N1}-\lambda_{\rm K})$, and the
dotted line through the K-band point is for $\alpha=-3$. Objects with $\alpha$\,$\gsim -2$ are believed to possess dusty circumstellar material.}
\label{f4}
\end{figure*}

TIMMI-2 was used on the first two nights to obtain mid-IR spectra of $\sigma$\,Ori, IRAS\,05358$-$0238 and TX\,Ori. With a slit of $3^{\prime\prime}$ wide and $50^{\prime\prime}$ long, the spectral resolving power, limited by the pixel scale of 0.02\,$\mu$m pixel$^{-1}$, was $R\sim200$--300 across a useful window of $\lambda=8$ to 13\,$\mu$m --- except for the spectral region between $\lambda=9$ and 9.9\,$\mu$m which was rendered useless due to a defunct channel in the TIMMI-2 array. The spectra were extracted from the frames resulting from the on-line data reduction pipeline using routines within {\sc midas}. Removal of telluric features (which were used to calibrate the wavelength axis) and absolute flux-calibration were achieved to an accuracy $\sim$\,30\% by comparison against HD\,4128 and HD\,32887 as well as the N1-band photometry.

\section{Results}

\subsection{Spectral energy distributions}

The shape of the IR spectral energy distribution (SED) is a very good indicator of the presence of circumstellar dust around an object and it can give some clues on the physical properties of the emitting material. Fig.\,\ref{f4} shows the SEDs of the target objects, ordered in luminosity (increasing from left to right and from top to bottom). The displayed photometry are 2MASS $J, H$ and $K_{\rm s}$ magnitudes, N1-band magnitudes or upper limits, and L$'$- and Q1-band magnitudes when available. The SEDs of $\sigma$\,Ori\,AB and $\sigma$\,Ori\,IRS\,1 have already been analysed elsewhere (van Loon \& Oliveira \cite{loon03}).

A convenient way of characterizing these SEDs is using a spectral index $\alpha$ (e.g.\, Kenyon \& Hartmann \cite{kenyon95}), defined as $\alpha = d\log (\lambda F_{\lambda}) / d \log \lambda$. According to this definition PMS stars are classified as Class I sources if $\alpha$\,$\gsim$\,0, Class II if  0\,$\gsim$\,$\alpha$\,$\gsim$\,$-2$ and Class III if $\alpha$\,$\sim$\,$-3$. In other words, Class III sources have IR colours consistent with normal stellar photospheres, Class II sources have IR excesses attributed to the presence of a circumstellar disk, and Class I sources have very large IR excesses normally associated with embedded PMS objects. In the case of our data, the spectral index is calculated as $\alpha=(F_{\rm N1}-F_{\rm K})/(\lambda_{\rm N1}-\lambda_{\rm K})$, indicated on the upper right corner for each object in Fig.\,\ref{f4}. The dotted line, the $\alpha$\,=\,$-$3 slope through the K-band point indicates the expected flux for a normal stellar photosphere. The top three objects, $\sigma$\,Ori\,E, $\sigma$\,Ori\,D and RX\,J05396$-$0242 show no excess at these wavelengths; most of the remaining objects show $\alpha$\,$> \,-$2 indicating the presence of cold dust, with the exception of the upper-limits for Haro\,5-26 and Kiso A-0976\,351.

\subsection{N-band spectroscopy}

The presence of the 10\,$\mu$m silicate feature not only unambiguously proves the existence of circumstellar dust, but its shape also offers clues on the mineralogy and geometry of the silicate grains. Silicate grains have been found in different environments ranging from molecular clouds to comets, and their detailed structure and composition trace the processing history of the circumstellar material.

Young stellar objects like T\,Tauri and Herbig Ae/Be stars exhibit a variety of silicate features, mostly in emission but in some cases in absorption (e.g.\, Hanner et al.\, \cite{hanner98}). We took N-band spectra of the brightest objects in our sample. The spectrum of the objects $\sigma$\,Ori\,AB and IRS1 helped to understand the nature of the IRAS emission associated with the $\sigma$\,Ori system and supported the discovery of the intriguing $\sigma$\,Ori\,IRS1 source (van Loon \& Oliveira \cite{loon03}).

\begin{figure}[tb]
\resizebox{\hsize}{!}{\includegraphics{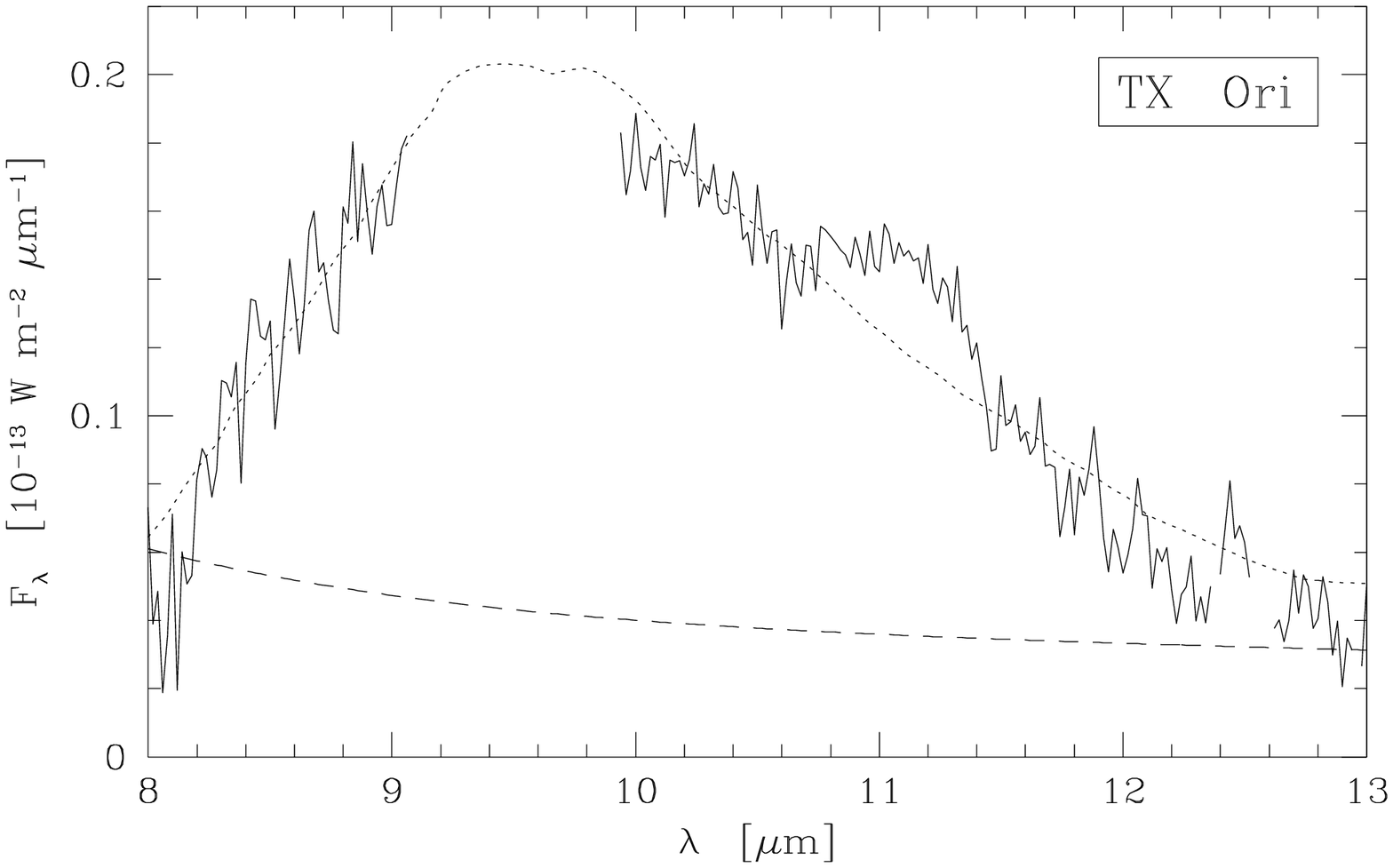}}
\caption{N-band spectrum of TX\,Ori. Using the procedure described in the text, the 10\,$\mu$m feature is shown to closely resemble the so-called ``Trapezium emissivity'' (dotted line), i.e\, it is readily identified as a silicate emission feature. There is, however, an extra feature at about 11.2\,$\mu$m. The dashed line is an estimate of the local continuum, as derived from the analysis of the SED (see section 4.3).}
\label{f5}
\end{figure}

\begin{figure}[tb]
\resizebox{\hsize}{!}{\includegraphics{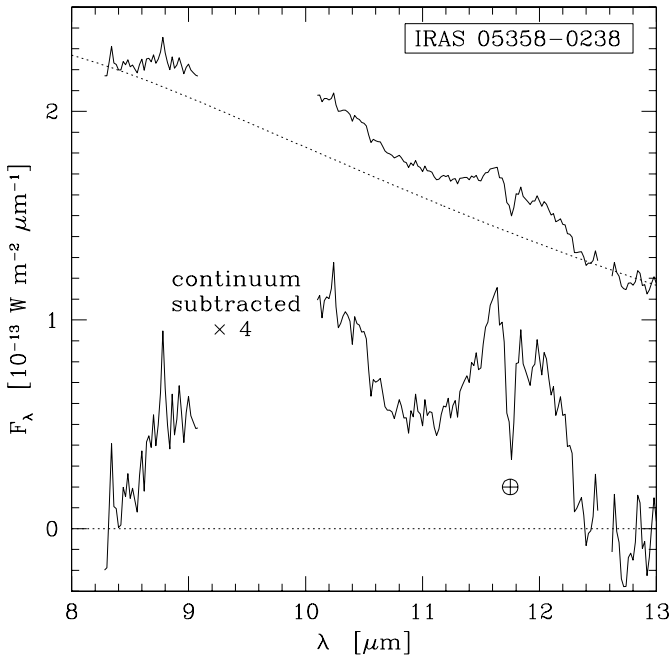}}
\caption{N-band spectrum of IRAS\,05358$-$0238. The dotted line on the top panel is an estimate of the local continuum, as derived from the analysis of the SED (see section 4.4). The bottom panel shows the continuum-subtracted spectrum, which is highly unusual.}
\label{f6}
\end{figure}

Figs.\,\ref{f5} and \ref{f6} show the N-band spectra of TX\,Ori and IRAS\,05358$-$0238 respectively. The spectrum of TX\,Ori clearly shows a typical broad silicate feature, peaking in the range 9$-$10\,$\mu$m, but with a smaller peak at about 11.2\,$\mu$m. On the other hand, the spectrum of IRAS\,05358$-$0238 is extremely unusual (see Section 4.2). 

\section{Discussion}

In this section we discuss the properties of the objects in our sample; TX\,Ori and IRAS\,05358$-$0238 are discussed in more detail in separate subsections.

\subsection{Early-type and Class\,III objects}

The near- and mid-IR SEDs of $\sigma$\,Ori E and D exhibit $\alpha \sim\,-3$
consistent with the slope of the Rayleigh-Jeans tail of a (relatively) hot blackbody distribution. This indicates the absence of significant dust material around these objects, as it is also expected from their early spectral types. RX\,J0539.6$-$0242 had been identified by Alcal\'{a} et al.\ (\cite{alcala96}) as a likely weak-line T\,Tauri star, based on optical spectrocopic criteria. The SED of RX\,J0539.6$-$0242 (with $\alpha \sim\,-3$ and no near-or mid-IR excess) implies the absence of a dusty circumstellar disk.

\subsection{Class\,II objects}

\begin{figure}[tb]
\resizebox{\hsize}{!}{\includegraphics{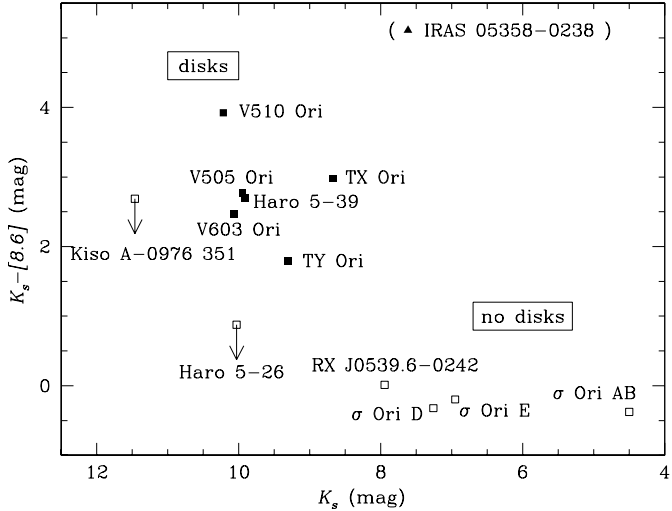}}
\caption{The $(K_{\rm s}-[8.6])$ vs.\ $K_{\rm s}$ colour-magnitude diagram for stars with (filled symbols) and without (open symbols) disks. There is a clear separation between brighter diskless objects and fainter objects with evidence of circumstellar material.}
\label{f7}
\end{figure}

TX\,Ori, TY\,Ori, Haro\,5-39, V510\,Ori, V505\,Ori and V603\,Ori have $\alpha \sim\,-1.6\,$ to $-0.2$, and are therefore classified as Class\,II sources, i.e\, objects with circumstellar disks. Fig.\,\ref{f7} illustrates the separation between two types of objects in our sample: brighter diskless objects and fainter, late(r)-spectral type objects with clear evidence of circumstellar disks. Based on this figure, it is unlikely that Haro\,5-26 possesses a significant circumstellar disk.

As mentioned previously, Haro\,5-39, V510\,Ori and V603\,Ori are the sources of the three well-known irradiated protostellar jets HH\,447, HH\,444 and HH\,445 respectively (Reipurth et al.\, \cite{reipurth98}). Our observations provide the first direct evidence of the presence of dust disks in these systems.

\begin{figure}[tb]
\resizebox{\hsize}{!}{\includegraphics{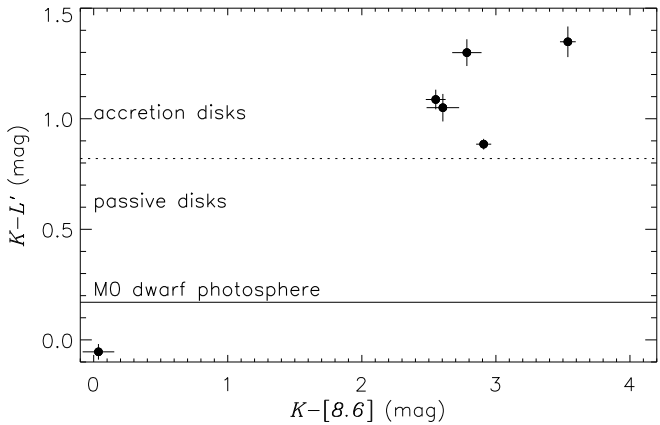}}
\caption{$KL'N1$ colour-colour diagram. The solid line corresponds to the $K-L^\prime$ colour of a main-sequence star with M0 spectral type. The dotted line separates passive from accretion disks, for a K7$-$M0 star, according to Wood et al.\ (\cite{wood02}).}
\label{f8}
\end{figure}

Fig.\,\ref{f8} shows the $KL'N1$ colour-colour diagram for the 6 objects with L$'$-band measurements. Again it is clear that all these objects have IR excesses, with the exception of RX\,J0539.6$-$0242. Even though with few data points and slightly different photometric bands, this diagram is qualitatively similar to diagrams described in the literature (e.g.\, Kenyon \& Hartmann \cite{kenyon95}; Haisch et al.\ \cite{haisch01b}). In particular, the Class\,II objects seem to clump together in the upper-right corner of the graph, suggesting a gap between Class\,II and Class\,III objects. This is in qualitative agreement with the predictions of the theoretical SEDs described by Wood et al.\ (\cite{wood02}): they find that the $K-L$ (and to a lesser extent $K-N$) signatures of circumstellar disks are largely insensitive to disk mass (over several orders of magnitude), decreasing very quickly for very low disk masses. Thus, from an observational point-of-view, one is likely to observe either objects with no $K-L$ excess (i.e.\, purely photospheric colours, no disks) or objects with significant excess, offering a possible explanation for the observed gap between the colours of Class\,III and Class\,II objects.

The Wood et al.\ (\cite{wood02}) theoretical models also suggest that large excesses in $K-L$ ($\gsim$\,0.7\,mag for a K7$-$M0 star) can only be achieved with a significant contribution from accretion luminosity in a massive disk. 
Indeed, for CTTS in Taurus-Auriga (Kenyon \& Hartmann \cite{kenyon95}), all objects with $(K-L) \gsim 1$ exhibit signatures of disk accretion (large H$\alpha$ equivalent width and UV continuum excess). In our sample, 4 of the 6 objects with L$'$-band measurements have large $K-L'$ excesses ($>1$\,mag), suggesting that these objects may possess massive actively accreting disks. This includes the three jet sources, where accretion of matter from a circumstellar disk is an important ingredient in the jet-collimation scenario.

\subsection{The T\,Tauri star TX\,Ori}

\begin{figure}[tb]
\resizebox{\hsize}{!}{\includegraphics{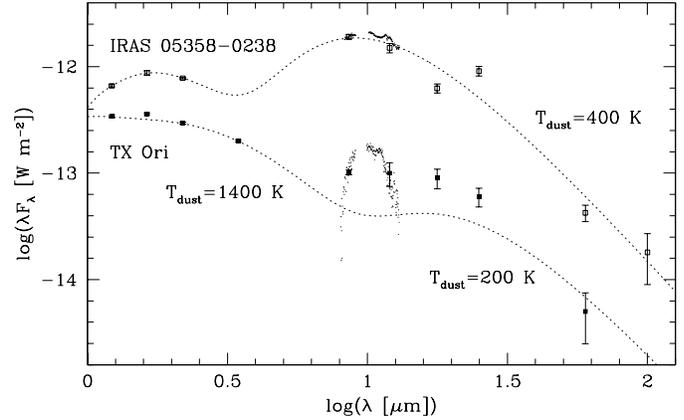}}
\caption{Spectral energy distributions of IRAS\,05358$-$0238 (open symbols)
and TX\,Ori (solid symbols). The available IR fluxes allow us to estimate a typical temperature of the dust in the circumstellar environments of the two objects: respectively 400\,K and 200\,K for IRAS\,05358$-$0238 and TX\,Ori. The K- and L-band fluxes of TX\,Ori require an additional warm dust component with a temperature $\sim$\,1400\,K. These simplified fits are used to estimate the continuum level in the region of the 10\,$\mu$m feature.}
\label{f9}
\end{figure}

With additional IRAS photometry and our mid-IR spectrum available for this object, we can attempt to derive some basic dust properties. Fig.\,\ref{f9} shows the IR SED of TX\,Ori (bottom, filled symbols). We started by decomposing the SED into two components: the photospheric contribution that dominates in the near-IR and a dust component responsible for the mid- and far-IR flux. The photospheric contribution is set to a blackbody with temperature $T_{\rm eff}$\,=\,4500\,K (in agreement with the object's spectral type). The excess IR emission was firstly represented by a single blackbody component with a temperature of $T_{\rm dust} \sim$\,200\,K. This temperature is uncertain, because we cannot rely on the Q1-band and 25\,$\mu$m flux densities as these are affected by another silicate feature at 18\,$\mu$m, which is likely to be strongly in emission too. As observed in other CTTS, this simple approach does not describe the observations in the near-IR (namely the K- and L-bands) and a significant contribution from a warmer dust component is needed to describe the SED. A typical temperature for this warm component would be $T_{\rm dust} \sim$\,1400\,K and may be attributed to accretion luminosity liberated in the disk (Wood et al.\ \cite{wood02}) or to scattered star-light. Although the superposition of three blackbody curves is a very crude attempt to ``model'' the SED, it does give insight into the global characteristics of the system and provides us with an estimate of the local continuum for the N-band spectrum.

As mentioned above, the analysis of the 10\,$\mu$m silicate feature provides useful insight into the chemical and physical properties of the circumstellar dust. Following the procedure described by Hanner et al.\ (\cite{hanner95,hanner98}, their case\,1), we first modelled the TX\,Ori N-band spectrum (Fig.\,\ref{f5}) by comparing with the so-called ``Trapezium emissivity'', which is believed to be the typical silicate feature of molecular cloud dust and which generally resembles the silicate feature in young stellar sources. This procedure provides a good fit to the TX\,Ori spectrum except for the emission feature at about 11.2\,$\mu$m. Several T\,Tauri stars and Herbig Ae/Be objects exhibit such feature, that cannot be understood solely in terms of an amorphous silicate emissivity. These have been identified as polycyclic aromatic hydrocarbons (PAHs) and/or crystalline silicate (forsterite). For both types of particles there are other emissivity peaks at other wavelengths but in both cases the peak at 11.2\,$\mu$m is the most conspicuous. Thus the distinction between these two contributions is not without difficulty, relying largely on the shape of the profile and the presence of other bands.

PAH emission features are usually expected in regions of high ultraviolet flux.
Accordingly, it has been identified in the IR spectra of many Herbig Ae/Be stars. It has been detected more rarely around luminous T\,Tauri stars (Hanner et al.\ \cite{hanner98}). Natta \& Kr\"{u}gel (\cite{natta95}) suggest that PAH emission can be produced around lower luminosity objects (e.g.\, an object of 1 L$_\odot$ and $T_{\rm eff}\simeq5000$\,K) but in many cases that feature will be swamped by the continuum emission of a circumstellar disk, and become undetectable. With a K4 spectral type, TX\,Ori seems too cool to excite such features.

Could PAH emission arise as a result of irradiation by $\sigma$\,Ori? Verstraete et al.\ \cite{verstraete01} describe the PAH emission spectra observed in NGC\,2023 and the Orion Bar, where molecular material is irradiated by a nearby hot star, creating especially favourable conditions for PAH molecules to abound and to be excited. We can scale the observed peak flux of the 11.2\,$\mu$m feature in these spectra to the case of TX\,Ori. The spectrum of TX\,Ori was obtained from a solid angle corresponding to an angular extent of about $1^{\prime\prime}$ (352 AU\footnote{For a Hipparcos distance of 352\,pc.}), and the radiation field of $\sigma$\,Ori is diluted over at least a projected distance of 0.89 pc. We then obtain an expected peak flux $\lambda F_\lambda < 8\times10^{-15}$\,W\,m$^{-2}$, i.e.\ at least one order of magnitude fainter than the 11.2\,$\mu$m feature observed in TX\,Ori. It seems thus unlikely that this feature is due to PAH molecules.

\begin{figure}[tb]
\resizebox{\hsize}{!}{\includegraphics{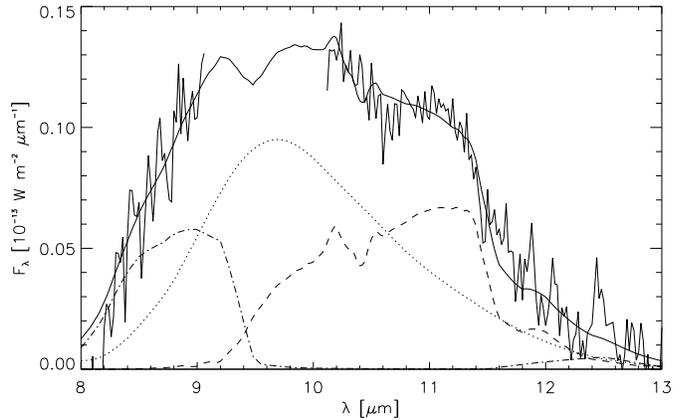}}
\caption{Fit to the continuum-subtracted spectrum of TX\,Ori. The solid line is the best-fit model spectrum, the dotted line is the (small grain) amorphous silicate contribution and the dashed and dashed-dotted lines are the contributions of the crystalline species, respectively forsterite and silica.}
\label{f10}
\end{figure}

On the other hand, crystalline olivines have also been detected in the spectra of many Herbig Ae/Be stars (e.g.\ Bouwman et al.\ \cite{bouwman01}) and a few T\,Tauri stars (Honda et al.\ \cite{honda03}; Meeus et al.\ \cite{meeus03}). In order to identify the different dust species that contribute to the 10\,$\mu$m feature, we follow the procedure described in Bouwman et al. (\cite{bouwman01}, see also Meeus et al.\, \cite{meeus03}). The continuum-subtracted spectrum of TX\,Ori (Fig.\,\ref{f10}) is fitted by a linear combination of several representative dust species: amorphous silicates (olivine, particle size 0.1 and 2\,$\mu$m to represent small and large dust grains), crystalline silicate (forsterite) and silica. All these species are found to be common in Herbig Ae/Be and CTTS disks. We are mostly interested in determining if forsterite could be responsible for the 11.2\,$\mu$m feature. Silica is included because a correlation was found between the amount of silica and the amount of forsterite for Herbig Ae/Be stars (Bouwman et al. \cite{bouwman01}). The presence of large silicate grains is an indication that dust processing in the form of dust particle coagulation has occurred. The adopted cross-sections can be found in Bouwman et al. (\cite{bouwman01}, and references therein).

Fig.\,\ref{f10} shows the best fit to the continuum-subtracted spectrum of TX\,Ori, as well as the contributions of each dust component. Clearly, attempting to fit the spectrum only with the small-grains amorphous silicate component would not work, forsterite and silica contribute significantly respectively between 10.5$-$11.5\,$\mu$m and 8$-$9\,$\mu$m. A good fit was also obtained when allowing for a component due to large amorphous silicate grains (but which still required similar amounts of forsterite and silica); as the inclusion of this component does not improve the quality of the fit, we opt to present the simplest fit. Thus, although we cannot draw conclusions on the presence of large grains, the shape of the emission feature can be reproduced by including a significant amount of crystalline dust --- in particular forsterite could be responsible for the 11.2\,$\mu$m feature. Therefore, we find evidence of dust processing in the form of crystallization (or annealing) but we do not find evidence for or against coagulation.

\subsection{The peculiar source IRAS\,05358$-$0238}

The near- and mid-IR SED of IRAS\,05358$-$0238 is characterised by $\alpha \sim +0.6$. In the context of young stellar objects, this would be classified as a Class\,I source, i.e.\ a substantial amount of warm dust enshrouds the central object. We also have 12--100\,$\mu$m IRAS flux densities for this object. Following the procedure described in the previous section we find that the IR contribution from the dust emission is well described by a single blackbody component with a temperature of $\sim$\,400\,K (Fig.\,\ref{f9}). The estimated ratio $L_{\rm IR}/L_{\star}$ is about 0.7, consistent with the object being a Class\,I object or a flat-spectrum source (Kenyon \& Hartmann \cite{kenyon95}).

The N-band spectrum of IRAS\,05358$-$0238 (Fig.\,\ref{f6}) does not show the broad silicate emission feature observed in most PMS objects with dusty circumstellar environments. Indeed, the spectrum is quite flat. The spectral slope in the interval 8.2$-$13\,$\mu$m is similar to what is observed for FU\,Orionis (Hanner et al.\ \cite{hanner98}), with a very weak emission feature around 10\,$\mu$m. This type of spectrum can be interpreted in terms of a
circumstellar disk which is optically thick at 10\,$\mu$m. An important
difference with FU\,Orionis, however, is the presence of another broad
emission feature in the spectrum of IRAS\,05358$-$0238, peaking between
11.6$-$11.8\,$\mu$m. We have been unable to identify a dust species that could
be responsible for this feature. In fact, nothing like this has ever been
observed in the mid-IR spectra of PMS objects or indeed of any object.

\begin{figure}[tb]
\resizebox{\hsize}{!}{\includegraphics{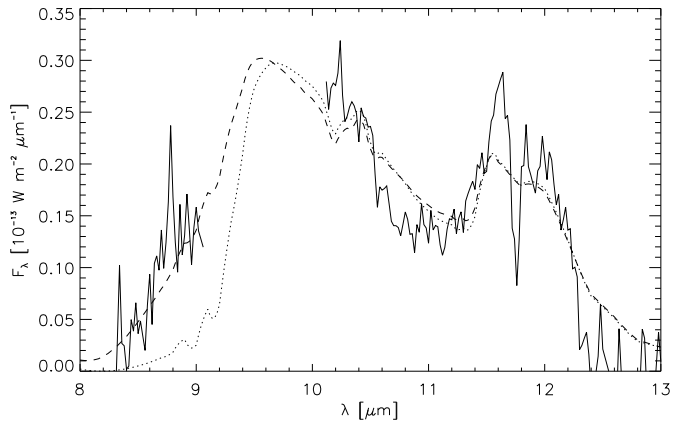}}
\caption{Fit to the continuum-subtracted spectrum of IRAS\,05358$-$0238. The dotted line represents a self-absorbed forsterite profile, while the dashed line further includes a contribution from small grains of amorphous silicate (see text).}
\label{f11}
\end{figure}

However, an acceptable --- though by no means perfect --- fit to the mid-IR
spectrum of IRAS\,05358$-$0238 is obtained if we allow forsterite to turn into
self-absorption (Fig.\,\ref{f11}). This alone reproduces the observed spectrum quite well in a qualitative sense, with optically thin emission in the wings peaking around 10 and 11.7\,$\mu$m, separated by a local minimum where the core of the feature goes into self-absorption around a wavelength of 11.3\,$\mu$m. The synthetic profile deviates from the observed feature in detail, but this may be due to differences in the exact composition, geometry and environmental
conditions of the dust grains (e.g.\, J\"{a}ger et al.\ \cite{jager98}). Note that the shape of the absorption dip that we observed in IRAS\,05358$-$0238 is somewhat rounder and peaks at a slightly shorter wavelength of $\lambda \sim 11.0$\,$\mu$m compared to the laboratory forsterite profile that we used, which is also the case for the 11\,$\mu$m feature that we observed in TX\,Ori (Fig.\,\ref{f10}).

The fairly high optical depth in the silicate feature is consistent with the $\sim$\,5 magnitudes of visual extinction needed to reproduce the near-IR part of the SED, and with the fact that the IR luminosity from the dust approaches the stellar luminosity, indicating that most of the stellar light is reprocessed by the circumstellar dust before leaving the system. A small amount of amorphous
olivine is still needed to correctly reproduce the emission around 9\,$\mu$m.
It must also be remarked that it is impossible to make a statement about the
crystallinity of the dust that is responsible for the underlying continuum
emission --- which is relatively strong compared to the mid-IR emission
feature in this object.

In order to shed light on the nature of this object, we obtained an optical
spectrum of the near-IR counterpart of the mid-IR source. The optical spectrum
is described in Appendix\,A and it seems to be that of an M5--5.5 giant. Members of the $\sigma$\,Orionis cluster are expected to have gravities around $\log g \sim 4.$ (Zapatero Osorio et al.\ \cite{osorio02}, for a typical cluster age of 3$-$5\,Myr), i.e.\ in between field dwarfs and giants (e.g.\ Zapatero Osorio et al.\ \cite{osorio02}; Oliveira et al. \cite{oliveira04}). This, together with the fact that this object is very bright ($K = 7.6$\,mag), lead us to believe that it is not a 3$-$5\,Myr member of the $\sigma$\,Orionis cluster. Maybe this object is significantly younger than the bulk of the cluster population, but that would be difficult to explain.

Maybe IRAS\,05358$-$0238 is an evolved giant star rather than a young star. Thick dust shells occur around highly evolved Asymptotic Giant Branch (AGB) stars and red supergiants (RSGs) as a result of mass loss. The mid-IR spectra of AGB stars and RSGs often show the amorphous silicate feature at 10\,$\mu$m in emission, absorption or self-absorption (depending on the optical depth) and sometimes a certain fraction of crystalline dust is invoked in order to explain all the IR spectral features. But no mid-IR spectrum exists in the literature that resembles that of IRAS\,05358$-$0238: for an overview of IR spectra of evolved objects, see e.g.\ Sylvester, Skinner \& Barlow (\cite{sylvester98}), Sylvester (\cite{sylvester99}), Sylvester et al.\ (\cite{sylvester99b}), Speck et al.\ (\cite{speck00}). The dust-enshrouded phase is in itself of short duration compared to the entire post-main-sequence evolution. This, and the uniqueness of its mid-IR spectrum, make it highly unlikely (but not impossible) to have encountered by chance an object like IRAS\,05358$-$0238 precisely in the direction of the $\sigma$\,Orionis cluster, and we therefore hesitate to rule out that it might still be a PMS object in the vicinity of $\sigma$\,Ori.

\section{Conclusions} 

We have performed mid-IR observations of the components of the $\sigma$\,Ori multiple system and of several suspected young stars in the vicinity of this system. We used N1-band observations to unequivocally ascertain whether circumstellar material is present around these objects. As expected, the early-type members of the multiple system were all found to be devoid of circumstellar dust (or free-free emission). From the suspected 10 young late-type objects, one object shows IR magnitudes consistent with the stellar photosphere and 7 objects clearly show evidence for circumstellar dust material, of which one object (IRAS\,05358$-$0238) we classify as Class\,I --- i.e.\, it exhibits a very substantial excess at these wavelengths. For 2 other objects we could only obtain upper limits for their N1-band brightness.

A comparison of $KL'N1$ colours with Taurus-Auriga observations and model expectations suggests that at least 4 objects (not including IRAS\,05358$-$0238, for which we do not have an L$^\prime$-band measurement) possess rather massive circumstellar disks and seem to be actively accreting from them. Only one of these objects has been confirmed as a member of the $\sigma$\,Ori cluster; however the detection of circumstellar dusty material hints at a young age, and it seems unlikely that these are all interlopers coming from other sites of more recent star formation. If these objects are indeed cluster members, it would imply that, at an age of 3$-$5\,Myr (e.g.\, B\'{e}jar et al.\ \cite{bejar99}; Oliveira et al.\ \cite{oliveira02}; Oliveira et al.\ \cite{oliveira04}), the more massive late-type cluster members still have massive accretion disks. The presence of accretion disks seems to extend to lower masses (Zapatero Osorio et al.\ \cite{osorio02}; Oliveira et al.\ \cite{oliveira04}). Either these objects are younger than the bulk of (non-accreting) cluster members or accretion disks (can) survive relatively long.

Our detections of mid-IR excess emission provide the first evidence for the presence of circumstellar disks in the irradiated-jet sources Haro\,5-39/HH\,447, V510\,Ori/HH\,444 and V603\,Ori/HH\,445. This supports the belief that accretion disks feed and help collimate the fast polar outflows responsible for the Herbig-Haro structures.

For the brightest (in the N1-band) objects IRAS\,05358$-$0238 and TX\,Ori, we also performed imaging in the Q1-band and spectroscopy in the N-band. This allowed us to probe the physical and chemical conditions of their circumstellar environments. The analysis of the SEDs of these objects provided us with typical circumstellar dust temperatures.

The mid-IR spectrum of TX\,Ori reveals a typical silicate emission feature, but with an extra component at 11.2\,$\mu$m that can be reproduced by addition of optically thin emission from crystalline silicate (forsterite). This can be regarded as evidence of dust processing. We are unable to establish whether dust particle coagulation has occurred as well.

The spectrum of IRAS\,05358$-$0238 is extremely unusual. The ratio of IR to stellar luminosity approaches unity, indicating the presence of a substantial amount of dust. However, the spectrum does not show the typical silicate feature; in fact, we are only able to describe it as being dominated by forsterite in self-absorption, something which has never been observed before. The status of this object remains uncertain, though, without evidence for (nor against) either its association with the $\sigma$\,Ori cluster or its youth.

\appendix

\section{IRAS data revisited}

\begin{table*}
\caption{Re-measured IRAS photometry, in Jy.}
\label{t3}
\begin{tabular}{llcccc}
\hline\hline
\multicolumn{1}{c}{Object}& \multicolumn{1}{c}{IRAS PSC ID}&\multicolumn{1}{c}{F$_{12}$}&\multicolumn{1}{c}{F$_{25}$}&\multicolumn{1}{c}{F$_{60}$}&\multicolumn{1}{c}{F$_{100}$}\\
&&\multicolumn{1}{c}{Jy$\pm\sigma$}&\multicolumn{1}{c}{Jy$\pm\sigma$}&\multicolumn{1}{c}{Jy$\pm\sigma$}&\multicolumn{1}{c}{Jy$\pm\sigma$}\\
\hline
 $\sigma$\,Ori\,IRS1 &IRAS\,05362$-$0237&  4.5$\pm$0.2	 &\llap{1}5$\pm$2     &\llap{1}5$\pm$4	&\llap{1}5$\pm$4\\
	             &IRAS\,05358$-$0238&  6.0$\pm$0.6	 & 7.6$\pm$0.8 & 0.85$\pm$0.15	& 0.6$\pm$0.3\\
 TX\,Ori             &IRAS\,05360$-$0245&  0.4$\pm$0.1	 & 0.5$\pm$0.1 & 0.10$\pm$0.05  &\\
 V510\,Ori 	&     &	0.15$\pm$0.05       &  0.28$\pm$0.04    & 0.5$\pm$0.1            &                \\                
\hline
\end{tabular}
\end{table*}

\begin{figure*}[ht]
\includegraphics[width=18cm]{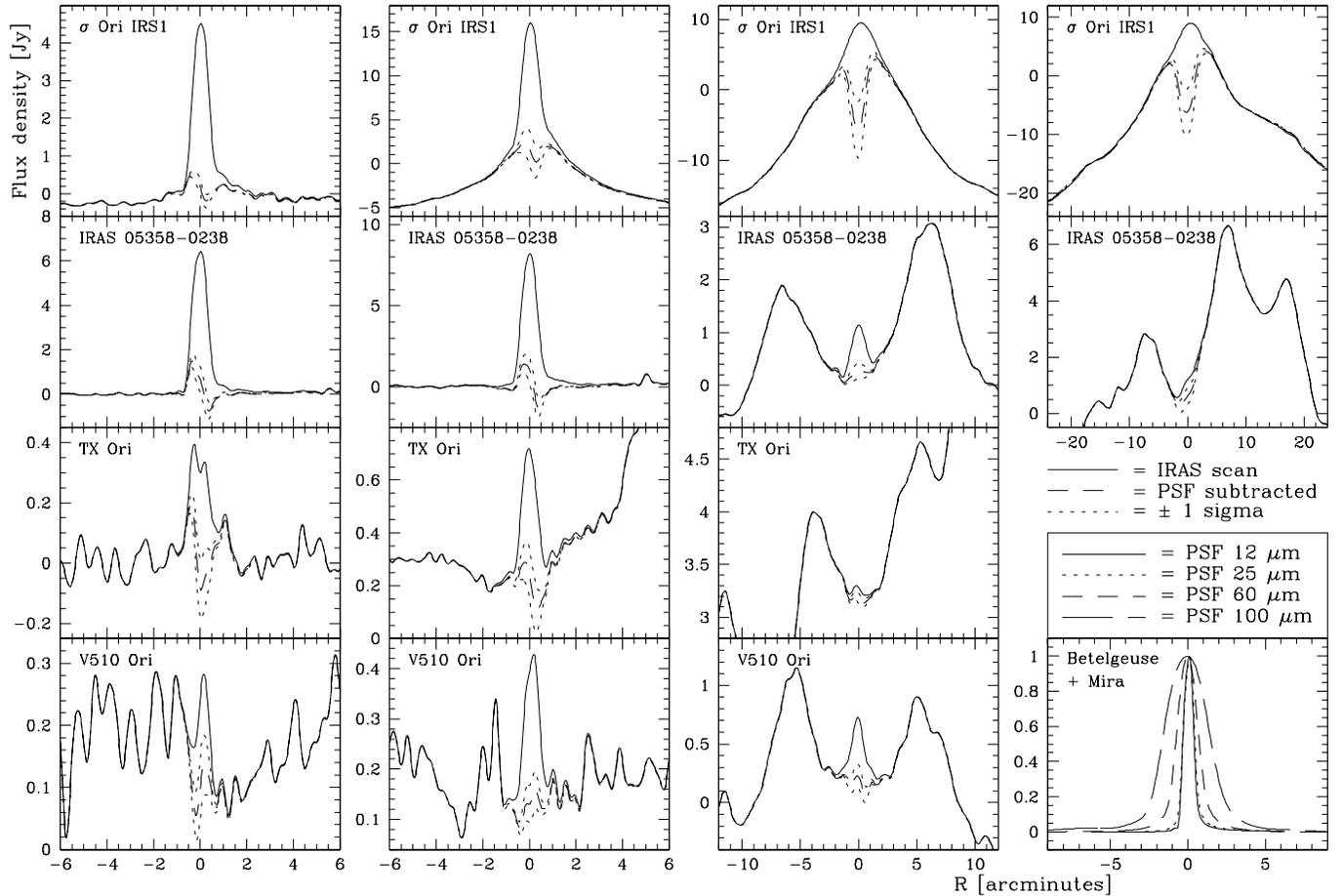}
\caption{Traces through the IRAS data obtained with Gipsy command {\sc scanaid}, centred on the positions of $\sigma$\,Ori\,IRS1, IRAS\,05358$-$0238, TX\,Ori and V501\,Ori (solid curves). Overplotted are the traces obtained when subtracting a PSF scaled to the estimated flux densities $F$ from Table\,A.1 (dashed curves) and when scaling to $F$+$\sigma_F$ or $F$$-$$\sigma_F$ (dotted curves). The PSFs are constructed by averaging the peak-normalised traces through Betelgeuse and Mira (lower right panel).}
\label{f12}
\end{figure*}

Data at 12, 25, 60 and 100\,$\mu$m were retrieved from the IRAS data base
server in Groningen (Assendorp et al.\ \cite{assendorp95}).
The Gipsy data analysis software was used to measure the flux density
from a trace through the position of the star (Gipsy command {\sc
scanaid}, see also Trams et al.\ \cite{trams99}). Sensible IRAS flux densities could be remeasured in this way only for $\sigma$\,Ori\,IRS1, IRAS\,05358$-$0238, TX\,Ori and V510\,Ori (Table\,A.1).

In Fig.\,\ref{f12} we show the IRAS scans for these sources. We overplot the traces after subtracting the Point Spread Function (PSF), scaled to the estimated flux densities listed in Table\,A.1. The PSF was constructed for each wavelength by averaging the scans through the positions of the bright IRAS sources Betelgeuse and Mira. The resulting PSFs are plotted in the lower right panel of Fig.\,\ref{f12}. Although some residuals in the PSF-subtracted scans remain, the source is usually removed quite well, lending credibility to the derived flux densities. Less certain measurements are obtained for $\sigma$\,Ori\,IRS1 at 60 and 100\,$\mu$m (which appears super-imposed on extended emission), IRAS\,05358$-$0238 at 100\,$\mu$m and TX\,Ori at 60\,$\mu$m, resulting in large errorbars on the photometry. The IRAS 12 and 25\,$\mu$m flux densities are in good agreement with our TIMMI-2 N1- and Q1-band measurements --- given the differences in passbands and the shapes of the SEDs.

\section{The optical spectrum of IRAS\,05358$-$0238}

An optical spectrum of IRAS\,05358$-$0238 was taken with EFOSC-2 at the ESO\,3.6\,m telescope at La Silla, Chile, during an instrument set-up night on 30 August 2003. Grism\,\#13 at a slitwidth of $1^{\prime\prime}$ with CCD\,\#40
provided a spectrum between $\lambda$\,=\,3680 and $\lambda$\,=\,9340\,\AA\ at a spectral resolution of about $\Delta\lambda\simeq25$\,\AA\ as measured from
the He-Ar calibration lamp spectra. Data reduction was done using standard
long-slit spectroscopy routines within the ESO software package {\sc midas}.
The response correction was obtained from a 10\,sec exposure of HD\,49798 at a
similar airmass (1.24 vs.\ 1.27) as the 1000\,sec exposure of IRAS\,05358$-$0238.

\begin{figure}
\resizebox{\hsize}{!}{\includegraphics{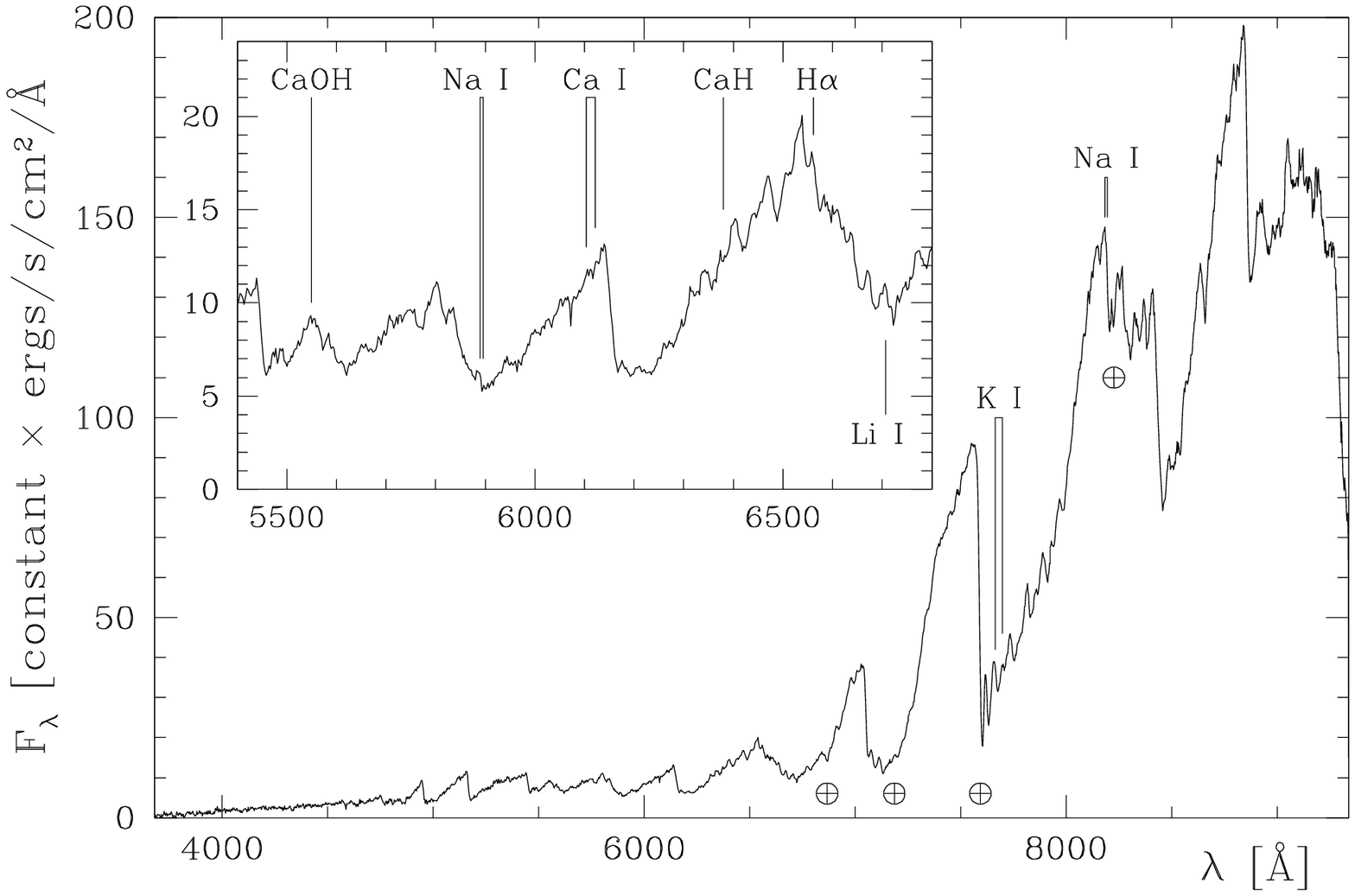}}
\caption{Optical spectrum of IRAS\,05358$-$0238. Several atomic and molecular features are identified. The inset shows the absence of CaOH and CaH, which suggests this object to be a giant.}
\label{f13}
\end{figure}

The spectrum of IRAS\,05358$-$0238 (Fig.\,\ref{f13}) is dominated by molecular
absorption bands of TiO, and a superficial comparison with spectral atlases
(e.g.\ Turnshek et al.\ \cite{turnshek85}) indicates a spectral type of M5$-$5.5. At such low photospheric temperature and at the present low spectral resolution, even the strongest atomic lines as indicated in Fig.\,\ref{f13} are only barely visible. It is thus impossible to, for instance, take the presence of conspicuous Li\,{\sc i} absorption at $\lambda$\,=\,6707\,\AA\ as evidence of young age. The absence of absorption bands of CaOH around $\lambda$\,=\,5550\,\AA\ and of CaH around $\lambda$\,=\,6380\,\AA\ does, however, provide strong evidence against high surface gravity, suggesting it is a giant rather than a dwarf. This spectral classification is confirmed by measurements of the optical colour ratios as defined by Kirkpatrick et al.\ (1991).

\begin{acknowledgements}
We would like to thank: Dr.\ Michael Sterzik for his support at the telescope; Dr.\ Vanessa Doublier for help with data reduction; and Drs.\ Emanuela Pompei and Lisa Germany and the ESO 3.6-m team for taking the optical spectrum of IRAS\,05358$-$0238 during an EFOSC-2 set-up night. We thank Dr.\ Jeroen Bouwman for providing us with the dust species cross-sections. We thank the anonymous referee for his/her interesting suggestions. This publication makes use of data products from the Two Micron All Sky Survey, which is a joint project of the University of Massachusetts and the Infrared Processing and Analysis Center/California Institute of Technology, funded by the National Aeronautics and Space Administration and the National Science Foundation. The IRAS data base server of the Space Research Organisation of the Netherlands (SRON) and the Dutch Expertise Centre for Astronomical Data Processing is funded by the Netherlands Organisation for Scientific Research (NWO). The IRAS data base server project was also partly funded through the Air Force Office of Scientific Research, grants AFOSR\,86-0140 and AFOSR\,89-0320. JMO acknowledges support of the UK Particle Physics and Astronomy Research Council.
\end{acknowledgements}


\begin{thebibliography}{}

\bibitem[1996]{alcala96}
Alcal\'{a} J.M., Terranegra L., Wichmann R. et al., 1996, A\&AS 119, 7

\bibitem[1995]{assendorp95}
Assendorp R., Bontekoe T.R., de Jonge A.R.W. et al., 1995, A\&AS 110, 395

\bibitem[1999]{bejar99}
B\'{e}jar V.J.S., Zapatero Osorio M.R., Rebolo R., 1999, ApJ 521, 671

\bibitem[1988]{bessell88}
Bessell M.S., Brett J.M., 1988, PASP 100, 1134

\bibitem[2000]{bodenheimer00}
Bodenheimer P., Hubickyj O., Lissauer J.J., 2000, Icar 143, 2

\bibitem[2001]{bouwman01}
Bouwman J., Meeus G., de Koter A. et al., 2001, A\&A 375, 950

\bibitem[2001]{carpenter01}
Carpenter J.M., 2001, AJ 121, 2851

\bibitem[1983]{castor83}
Castor J.I., Simon T., 1983, ApJ 265, 304

\bibitem[1990]{garcia90}
Garc\'{\i}a-Lario P., Manchado A., Suso S.R., Pottasch S.R., Olling R., 1990, A\&AS 82, 497

\bibitem[1994]{green94}
Green T.P., Wilking B.A., Andr\'{e} P., Young E.T., Lada C.J., 1994, ApJ 434, 614

\bibitem[1982]{groote82}
Groote D., Hunger K., 1982, A\&A 116, 64

\bibitem[1997]{gullbring97}
Gullbring E., Hartmann L., Brice C., Calvet N., 1997, ApJ 492, 323

\bibitem[1995]{hanner95}
Hanner M.S., Brooke T.Y., Tokunaga A.T., 1995, ApJ 438, 250

\bibitem[1998]{hanner98}
Hanner M.S., Brooke T.Y., Tokunaga A.T., 1998, ApJ 502, 871

\bibitem[2001]{haisch01a}
Haisch K.E., Lada E.A., Lada C.J., 2001a, ApJ 553, 153

\bibitem[2001]{haisch01b}
Haisch K.E., Lada E.A., Pi\~{n}a R.K, Telesco C.M., Lada C.J.,  2001b, AJ 121, 1512

\bibitem[1998]{hartmann98}
Hartmann L., Calvet N., Gullbring E., D'Alessio P., 1998, ApJ 495, 385

\bibitem[2003]{honda03}
Honda M., Kataza H., Okamoto Y.K., 2003, ApJ 585, 59

\bibitem[1998]{jager98}
J\"{a}ger C., Molster F.J., Dorschner J. et al., 1998, A\&A 339, 904

\bibitem[1995]{kenyon95}
Kenyon S.J., Hartmann L., 1995, ApJS 101, 117

\bibitem[1991]{kirkpatrick91}
Kirkpatrick J.D., Henry T.J., McCarthy D.W., 1991, ApJS 77, 417

\bibitem[2003]{meeus03}
Meeus G., Sterzik M., Bouwman J., Natta A., 2003, A\&A 409, L25

\bibitem[1997]{meyer97}
Meyer M.R., Calver N., Hillenbrand L.A., 1997, AJ 114, 288

\bibitem[1995]{natta95}
Natta A., Kr\"{u}gel E., 1995, A\&A 302, 849

\bibitem[2003]{nurnberger03}
N\"{u}rnberger D.E.A., Stamke T.,  2003, A\&A 400, 223

\bibitem[2002]{oliveira02}
Oliveira J.M., Jeffries R.D., Kenyon M.J., Thompson S.A., Naylor T., 2002, A\&A 382, 22

\bibitem[2004]{oliveira04}
Oliveira J.M., Jeffries R.D., van Loon J.Th., 2004, MNRAS 347, 1327

\bibitem[1998]{reipurth98}
Reipurth B., Bally J., Fesen R.A., Devine D., 1998, Nature 396, 343

\bibitem[1985]{rieke85}
Rieke G.H., Lebofsky M.J., 1985, ApJ 288, 618

\bibitem[2000]{speck00}
Speck A.K., Barlow M.J., Sylvester R.J., Hofmeister A.M., 2000, A\&AS 146, 437

\bibitem[1999]{sylvester99}
Sylvester R.J., 1999, MNRAS 309, 180

\bibitem[1998]{sylvester98}
Sylvester R.J., Skinner C.J., Barlow M.J., 1998, MNRAS 301, 1083

\bibitem[1999]{sylvester99b}
Sylvester R.J., Kemper F., Barlow M.J. et al., 1999, A\&A 352, 587

\bibitem[1999]{trams99}
Trams N.R., van Loon J.Th., Waters L.B.F.M. et al., 1999, A\&A 346, 843

\bibitem[1985]{turnshek85}
Turnshek D.E., Turnshek D.A., Craine E.R., Boeshaar P.C., 1985, An atlas
of digital spectra of cool stars. Western Research Company, Tucson

\bibitem[2003]{loon03}
van Loon J.Th., Oliveira J.M., 2003, A\&A 405, L33

\bibitem[2001]{verstraete01}
Verstraete L., Pech C., Moutoun C. et al., 2001, A\&A 372, 981

\bibitem[1997]{walter97}
Walter F.M., Wolk S.J., Freyberg M., Schmitt J.H.M.M., 1997, MmSAI 68, 1081

\bibitem[1992]{weaver92}
Weaver W.B., Jones G., 1992, ApJS 78, 239

\bibitem[2002]{wood02}
Wood K., Lada C.J., Bjorkman J.E. et al., 2002, ApJ 567, 1183

\bibitem[1996]{wolk96}
Wolk S.J., 1996, Ph.D. thesis, State Univ. New York at Stony Brook

\bibitem[2002]{osorio02}
Zapatero Osorio M.R., B\'{e}jar V.J.S., Pavlenko Y. et al., 2002,
A\&A 384, 937


\end{thebibliography}
\end{document}